\documentclass[11pt,a4paper]{article}
\usepackage{latexsym}
\usepackage{cite}
\usepackage{epsfig}
\usepackage{amssymb}
\usepackage{latexsym}
\usepackage{epsfig, subfigure, subfloat, graphicx, float}
\usepackage{mathrsfs}
\usepackage{amsmath}
\usepackage{color}
\usepackage[usenames]{xcolor}
\graphicspath{{Image/}}
\author{H. Mohseni Sadjadi\footnote{mohsenisad@ut.ac.ir}  and V. Anari\footnote{v.anari@ut.ac.ir}
\\ {\small Department of Physics, University of Tehran,}
\\ {\small P. O. B. 14395-547, Tehran 14399-55961, Iran}}
\title{Cosmic acceleration and de Sitter expansion in hybrid mass varying neutrino model }
\begin{document}
\maketitle
\begin{abstract}
A nonminimally coupled hybrid dark energy model is introduced. The dark energy evolution is triggered by mass varying neutrinos. Quintessence evolution begins from an initial point, with an insignificant dark energy density, and arrives at a final de Sitter attractor. While initially, for a Higgs-like potential, the non-minimal coupling keeps the system in a state with negligible dark energy density, finally the second quintessence conducts the evolution to a stable fixed point. These evolutions occur through successive $Z_2$ symmetry breakings.
\end{abstract}

\section{Introduction}

 There is much evidence that the current expansion of the Universe is accelerating \cite{acc1,acc2,acc3,acc4}. To describe the origin of this phenomenon, many attempts such as introducing additional dark energy hypothetical fields, like the quintessence scalar field, modifying the Einstein theory of gravity and so on,  have been made \cite{dark1,dark2,dark3,dark4,dark5,dark6,dark7,dark8,dark9,dark10,dark11,dark12,dark13,dark14,dark15,dark16,dark17,dark18}. Despite the negative equation of state parameter (EoS) of dark energy, dark energy and dark matter densities are of the same order today. Hence the dark energy density has been negligible in the past and earlier epochs and only in the recent time it has become dominant.  To describe this via a dynamical dark energy, one may assume that something happened to activate the dark energy component. This activation could be realized by interactions between dark energy and other components such that by the evolution of these components the dark energy behavior changed.  As an example assume that, due to the interaction of the quintessence with the matter, the quintessence squared effective mass becomes $\mu_{eff.}^2=\rho_m-\rho_c$. When, due to the redshift, the matter density ($\rho_m$) becomes less than the critical density ($\rho_c$), we have $\mu_{eff.}^2<0$. For a Higgs like potential, this triggers the evolution of the quintessence. This idea was proposed and employed in the symmetron cosmology model \cite{kh,sym,ka}, and is based on $Z_2$ symmetry breaking. However, in the symmetron model, through the symmetry breaking, the quintessence potential becomes negative unless we add a cosmological constant to the theory to provide the positive potential required for the acceleration \cite{sad1}. This makes this theory indistinguishable from $\Lambda CDM$ \cite{kh}. In this model, like some other screening models, the coupling may also be derived from a conformal coupling in matter sector, resulting in a term proportional to $\rho_m-3P_m$ in the quintessence squared effective mass term.  For pressureless matter, we have $P_m=0$, and the quintessence behavior is only affected by the matter density $\rho_m-3P_m=\rho_m$.

To activate the quintessence through the symmetry breaking, there are also other possibilities. For example consider a component, denoted by $\nu$, which was relativistic in the early epoch: $\rho_\nu-3P_\nu=0$,  but became non-relativistic later, so that $\rho_\nu-3P_\nu=\rho_\nu$, changing the quintessence effective potential at the late time. This can force dark energy to start its evolution through the $Z_2$ symmetry breaking. Inspired by \cite{fardon}, this idea was proposed in \cite{anari}, where the rise of dark energy was related to the massive neutrinos. However, in that model, the massive neutrinos redshift like dust, and the effective potential, affected by $\rho_\nu$, reduces finally to its initial form and the acceleration is not persistent. To have a permanent acceleration like a de Sitter expansion at the late time, an additional coupling is required. For example, the symmetron model with a non-minimal coupling to the torsion scalar has a late time de Sitter stable solution but with the price of losing the Lorentz symmetry \cite{sad2,sad2.1}.  A more natural assumption is the non-minimal coupling of the quintessence $\phi$ to the Ricci scalar through a term proportional to $R\phi^2$,  which can be justified by renormalizability in first loop corrections. This model has been employed in the literature to study the inflation and the late time acceleration \cite{faraoni}. Equipped with a hybrid quintessence, this model has a de Sitter attractor solution \cite{sad3}.

In this paper, we try to describe the onset of the dark energy by $Z_2$ symmetry breaking caused by massive neutrinos. Using the nonminimally coupled hybrid model, we investigate the possibility to have an attractor solution corresponding to the late time de Sitter expansion.

The scheme of the paper is as follows: In the second section, we non-minimally couple the quintessence to the Ricci scalar and also to the massive neutrinos and show whether the change of the neutrino behavior, from relativistic to non-relativistic, can trigger the evolution of the quintessence through the $Z_2$ symmetry breaking. In the third section, we investigate the possibility to have an attractor solution in the corresponding hybrid model. We finally conclude our results in the fourth section.

Throughout the paper we use the natural units with $8\pi G=1$, hence  $\hbar=c=8\pi G=1$. We take also the Boltzmann constant as $k_B=1$ and our convention for the metric signature is (-,+,+,+).

\section{Mass varying neutrino in non-minimally coupled quintessence model, and cosmic acceleration}
We begin with the action
\begin{eqnarray}\label{action}
S&=&\int d^4x\sqrt{-g}\left[ \frac{R}{2}-\frac{1}{2}(\partial_{\mu}\phi\partial^{\mu}\phi +\epsilon R\phi^2) -V(\phi)\right]\nonumber
\\& &+S_\nu[\tilde{g}_{\mu\nu}]+S_m+S_r
\end{eqnarray}
where $r$ stands for the relativistic matter, $m$ for the pressureless matter and $\nu$ for the neutrinos. Note that in our units, the reduced Planck mass is $M_P=1$.  $V(\phi)$ is the quintessence potential. $\epsilon\in \Re$, and $R$ is the Ricci scalar. There is a conformal coupling in the neutrino sector $\tilde{g}_{\mu\nu}=A^2(\phi)g_{\mu\nu}$, where $A(\phi)$ is a positive function. Note that the quintessence-neutrino interaction could be also introduced in the context of mass varying neutrino without using the conformal coupling method \cite{fardon,anari}.
By adopting a spatially flat Friedmann–-Lemaitre–-Robertson–-Walker (FLRW) metric
\begin{equation}
ds^2=-dt^2+a^2(t)\left(dx^2+dy^2+dz^2\right),
\end{equation}
where $a(t)$ is the scale factor, variation of (\ref{action}) with respect to the metric yields to the Friedmann equation
\begin{equation}\label{fried1}
H^2=\frac{1}{3}\sum_i\rho_i,\,\, (i=r, m, \nu , \phi)
\end{equation}
with $H=\dfrac{\dot{a}}{a}$ being the Hubble parameter, and the evolution of the Hubble parameter is given by
\begin{equation}\label{Hdot1}
\dot{H}=-\frac{1}{2}\sum_i(\rho_i + P_i),
\end{equation}
where $\rho_i$ and $P_i$ are the energy density and pressure of the ith cosmic fluid, respectively. Also, the energy density $\rho_\phi$ and pressure $P_\phi$ of the quintessence are given by
\begin{equation}\label{rho_phi}
\rho_\phi = \frac{1}{2}\dot{\phi}^2+V(\phi)+\epsilon \left( 6H\phi\dot{\phi}+3H^2\phi ^2 \right),
\end{equation}
and
\begin{equation}\label{P_phi}
P_\phi =\frac{1}{2}\dot{\phi}^2-V(\phi)-\epsilon\left( 4H\phi\dot{\phi}+2\dot{\phi}^2+2\phi\ddot{\phi}+\left(2\dot{H}+3H^2\right)\phi^2 \right).
\end{equation}
For cosmic acceleration we require to have $H^2+\dot{H}>0$. For $\dot{H}>0$ we have super-acceleration which implies $P_\phi<-\rho_\phi$. Note that in the case of minimal coupling, i.e. $\epsilon=0$, we have $P_\phi =\frac{1}{2}\dot{\phi}^2-V(\phi)$ and $\rho_\phi = \frac{1}{2}\dot{\phi}^2+V(\phi)$, and the super-acceleration is forbidden.

Variation of (\ref{action}) with respect to $\phi$ gives
\begin{equation}\label{EoMphi}
\ddot{\phi}+3H\dot{\phi}+V_{eff.,\phi}(\phi)=0 ,
\end{equation}
where $V_{eff.,\phi}(\phi)=\dfrac{dV_{eff.}(\phi)}{d\phi}$ and the effective potential is given by
\begin{equation}\label{V_eff}
V_{eff.,\phi}(\phi)=\epsilon R\phi + V_{,\phi} + A_{,\phi}A^{-1}(\phi)(\rho_\nu - 3P_\nu) .
\end{equation}
where $R$ is the Ricci scalar which in terms of the Hubble parameter is $R=6\dot{H}+12H^2$. Also, the continuity equations are given by
\begin{equation}\label{CEm}
\dot{\rho}_m+3H\rho_m=0,
\end{equation}
\begin{equation}\label{CEr}
\dot{\rho}_r+4H\rho_r=0,
\end{equation}
\begin{equation}\label{CEnu}
\dot{\rho}_\nu +3H(\rho_\nu +P_\nu)=A_{,\phi}A^{-1}\dot{\phi}(\rho_\nu - 3P_\nu).
\end{equation}

By using Fermi-Dirac distribution for neutrinos which are in thermal equilibrium and whose masses are $\phi$ dependent, one finds \cite{Brookfield, Peccei}
\begin{equation}\label{CEmvnu}
\dot{\rho}_\nu +3H(\rho_\nu +P_\nu)=\frac{m_{\nu,\phi}(\phi)}{m_\nu(\phi)}\dot{\phi}(\rho_\nu - 3P_\nu),
\end{equation}
therefore, taking $A(\phi)=\frac{m_\nu(\phi)}{M}$ in (\ref{CEnu}),  results in (\ref{CEmvnu}), where $M$ is a mass scale. In other words the two approaches, i.e. mass varying neutrino whose mass depends on the scalar field $\phi$ \cite{Pettorino, Brookfield, Peccei}, and conformal coupling in neutrino sector through a function of $\phi$ \cite{anari} , give the same equations of motion \cite{anari}.

For relativistic neutrinos, $m_\nu \ll T_\nu$, we have
\begin{eqnarray}\label{r_eq}
&&\ddot{\phi}+3H\dot{\phi}+\epsilon R\phi + V_{,\phi} = 0 \nonumber \\
&&\dot{\rho}_\nu +4H\rho_\nu =0,
\end{eqnarray}
and $V_{eff.,\phi}=\epsilon R\phi + V_{,\phi}$. While for non-relativistic neutrinos, $m_\nu \gg T_\nu$,
\begin{eqnarray}\label{nr_eq}
&&\ddot{\phi}+3H\dot{\phi}+\epsilon R\phi +V_{,\phi}+A_{,\phi}A^{-1}\rho_\nu =0 \nonumber \\
&&\dot{\rho}_\nu +3H\rho_\nu =A_{,\phi}A^{-1}\dot{\phi}\rho_\nu .
\end{eqnarray}
In this case, we can define a rescaled energy density $\hat{\rho}_\nu=A^{-1}\rho_\nu$ in terms of which (\ref{nr_eq}) reduces to
\begin{equation}\label{CEnu2}
\dot{\hat{\rho}}_\nu+3H\hat{\rho}_\nu=0 ,
\end{equation}
and
\begin{equation}\label{EoMphi2}
\ddot{\phi}+3H\dot{\phi}+V_{eff.,\phi}(\phi)=0 ,
\end{equation}
where $V_{eff.,\phi}=\epsilon R\phi +V_{,\phi}+A_{,\phi}\hat{\rho}_\nu$.\\

To construct our model we proceed as follows: We require that both the  potential, $V(\phi)$,  and $A(\phi)$ have $Z_2$ symmetry.  We assume that the scalar field is initially at the stable point $\phi=0$, where $V(\phi)=0$, until the $Z_2$ symmetry breaks. Before the symmetry breaking  $\phi=0$  and $\dot{\phi}=0$, so according to (\ref{rho_phi}): $\rho_\phi=0$  and the dark energy density is negligible.  The quintessence squared mass is  given by $\tilde{\mu}^2=\frac{d^2V}{d\phi^2}|_{\phi=0}$ and the squared effective mass is defined as
\begin{equation}
\mu^2_{eff.}=\frac{\partial^2 V_{eff.}}{{\partial \phi}^2}\Bigm|_{\phi=0},
\end{equation}
and according to (\ref{V_eff}), one finds
\begin{equation}\label{mu_eff}
\mu^2_{eff.}=\tilde{\mu}^2+\epsilon\left(\rho_m+(\rho_\nu - 3P_\nu)\right)+A^{-1}(0)A_{,\phi\phi}\Bigm|_{\phi=0}(\rho_\nu - 3P_\nu).
\end{equation}
Initially, the neutrinos are relativistic ($\rho_\nu\approx3P_\nu$) and for $\phi=0$  to be a stable solution, it is necessary that the squared effective mass be positive which means
\begin{equation}\label{condition1}
\tilde{\mu}^2+\epsilon\rho_m >0
\end{equation}
For $\tilde{\mu}^2<0$, this leads to $\epsilon>0$. For $\tilde{\mu}^2>0$ and $\epsilon>0$, (\ref{condition1}) holds for all values of $\rho_m$.  In the following, we take $\epsilon>0$. A similar analysis may be carried out for $\epsilon<0$. In this initial state, $\phi$ does not contribute in the Universe evolution and obviously $\dot{H}+H^2<0$, and the Universe is decelerating. As the universe expands and temperature decreases, neutrinos exit from the relativistic state and $\rho_\nu-3P_\nu$ becomes relevant. We require that this gives rise to the $Z_2$ symmetry breaking. For $\phi=0$ to become an unstable point, the squared effective mass has to become negative
\begin{equation}\label{condition2}
\tilde{\mu}^2+\epsilon\rho_m+\left(\epsilon +A^{-1}(0)A_{,\phi\phi}\Bigm|_{\phi=0}\right)\left(\rho_\nu - 3P_\nu\right)<0
\end{equation}
As the temperature of the universe decreases, $\rho_\nu - 3P_\nu$ increases from zero such that (\ref{condition2}) holds, therefore the effective potential becomes concave at $\phi=0$ and the scalar field rolls down its effective potential, giving rise to a non-zero effective dark energy density. Note that this plan necessitates that $A_{,\phi\phi}\bigm|_{\phi=0}<0$. From this point, the evolution of the Universe is governed by the following set of equations:
\begin{eqnarray}\label{eq_set}
&&\dot{a}-Ha=0,\nonumber\\
&&\dot{\hat{\rho}}_\nu+3H\hat{\rho}_\nu=0,\nonumber\\
&&\dot{\rho}_m+3H\rho_m=0,\nonumber\\
&&\dot{\rho}_r+4H\rho_r=0,\nonumber\\
&&\ddot{\phi}+3H\dot{\phi}+\epsilon R\phi +V_{,\phi}+A_{,\phi}\hat{\rho}_\nu=0,\nonumber\\
&&-2\left(1+(6\epsilon^2-\epsilon)\phi^2\right)\dot{H}=(1-2\epsilon)\dot{\phi}^2+8\epsilon H\phi\dot{\phi}+24\epsilon^2 H^2\phi^2\nonumber \\
&&+2\epsilon\phi V_{,\phi} +2\epsilon\phi A_{,\phi}\hat{\rho}_\nu+\rho_m+\frac{4}{3}\rho_r+A(\phi)\hat{\rho}_\nu.
\end{eqnarray}
The last above equation is derived using (\ref{Hdot1}), (\ref{rho_phi}), (\ref{P_phi}) and (\ref{EoMphi}).  Solving the above equations analytically is very complicated if not impossible. So to show how the model works, we illustrate our results with a numerical example. In our numerical study, we ignore the neutrinos pressure because in the presence of $P_\nu$, (\ref{eq_set}) becomes very complicated to solve, even numerically. Also, like\cite{Brookfield, Pietroni, anari}, we choose $A(\phi)=e^{-\alpha\phi^2}$ which satisfies all the conditions mentioned, and where $\alpha$ is a positive constant.
We choose the Higgs like potential as
\begin{equation}\label{potential1}
V(\phi)=-\frac{1}{2}\mu^2\phi^2+\frac{1}{4}\lambda\phi^4 \,  , \hspace{2cm} (\lambda>0).
\end{equation}
and the parameters of the model as
\begin{equation}\label{para}
\epsilon=1, \hspace{1cm} \alpha=5, \hspace{1cm} \mu=4 \, H_0, \hspace{1cm} \lambda=10^{-4} \, H_0^2.
\end{equation}
We choose the following initial conditions
\begin{align}\label{ICs}
&a=\frac{1}{2000+1},\,\, H =6.77\times10^4 \, H_0,\,\, \phi =10^{-5},\,\, \dot{\phi}=10^{-4} \, H_0 , \nonumber \\
&\rho_r =4.38\times10^{9} \, H_0^2,\,\, \rho_m =7.02\times10^{9} \, H_0^2 ,\,\, \hat{\rho}_\nu =2.34\times10^{9} \, H_0^2,
\end{align}
where $H_0$ is the present Hubble parameter (the Hubble parameter at $a=1$).

We define dimensionless time $\tau$ as $\tau=tH_0$ and the initial conditions are set at $\tau=0$ when the redshift is $z=2000$, i.e. after matter-radiation equality.
$\tau=0$ is set when the neutrinos are completely non-relativistic, i.e. $\rho_\nu - 3P_\nu \simeq \rho_\nu$. It is worth noting that the scalar field began its motion just after symmetry breaking where $P_\nu$ was not completely negligible (at $\tau<0$).   We have chosen the parameters such that the $Z_2$ symmetry breaking is only due to the neutrinos.

The relative densities defined by $\Omega_i=\frac{\rho_i}{3H^2}$ are derived from (\ref{ICs}) as
\begin{equation}
\Omega_r =0.319,\,\, \Omega_m =0.511,\,\, \Omega_\nu =0.170,\,\, \Omega_\phi =10^{-10},
\end{equation}
according to $\Omega_\phi=10^{-10}$, we can see that the chosen initial values for $\phi$ and $\dot{\phi}$ give a negligible contribution of dark energy in the total density. Note that  ordinary neutrinos which have electroweak interactions, were in thermal equilibrium with ordinary species in the (very) early universe. Here, one can assume that the mass varying neutrinos do not have electroweak interaction (like sterile neutrinos), so the usual relations describing neutrino-photon density ratio may not be still valid.\\

The deceleration parameter $q=-\dfrac{a\ddot{a}}{\dot{a}^2}$, in terms of the Hubble parameter is given by
\begin{equation}
q=-\left(1+\frac{\dot{H}}{H^2}\right),
\end{equation}
where $q<0$ yields to a positively accelerated universe and in the minimal coupling models($\epsilon=0$), $-1<q<2$. While for the non-minimal coupling models($\epsilon\neq0$), $q<-1$ is possible, which is equivalent to super-acceleration.\\

\begin{figure}[H]
\centering
\includegraphics[width=6 cm]{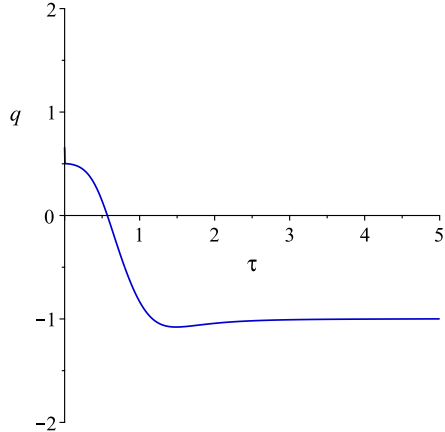}
\caption{The deceleration parameter in terms of $\tau$, for the parameters (\ref{para}) and initial conditions (\ref{ICs}).}
\label{fig_q}
\end{figure}

In fig.(\ref{fig_q}), the deceleration parameter $q$ is plotted in terms of $\tau$. As we can see, the universe is transited from a deceleration epoch to an acceleration epoch at a time less than the Hubble time. This acceleration begins at the redshift $z\simeq0.5$, and at redshift $z\simeq-0.2$, $q$ becomes slightly less than $-1$ which indicates super-acceleration of the universe.

\begin{figure}[H]
\centering
\includegraphics[width=6 cm]{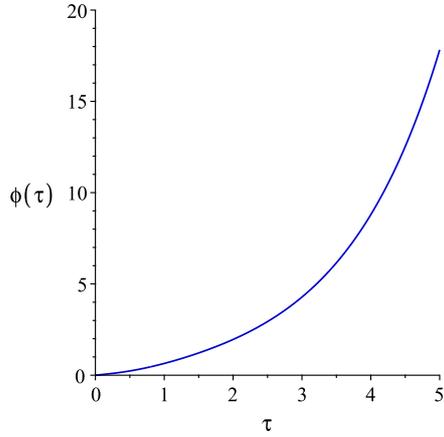}
\caption{The scalar field in terms of $\tau$, for the parameters (\ref{para}) and initial conditions (\ref{ICs}).}
\label{fig_phi}
\end{figure}

In fig.(\ref{fig_phi}) the scalar field grows from $\phi=10^{-5}$, but this growth does not stop and results in a singularity at $\tau\simeq10$.

\begin{figure}[H]
\centering
\includegraphics[width=10 cm]{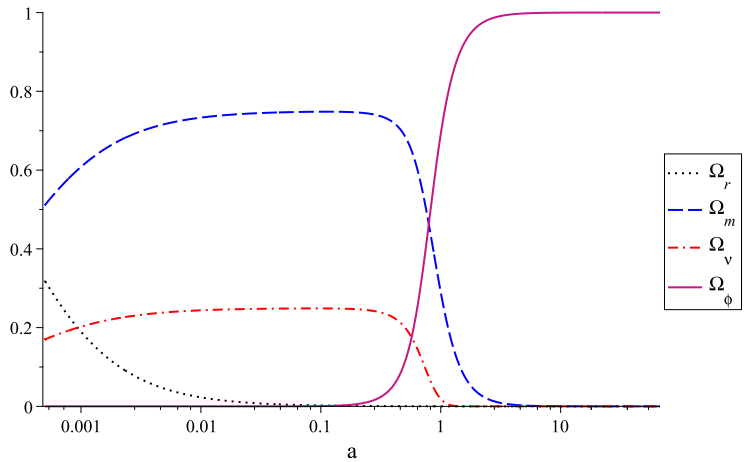}
\caption{Relative densities in terms of the scale factor $a$, for the parameters (\ref{para}) and initial conditions (\ref{ICs}).}
\label{fig_Omega}
\end{figure}

In fig.(\ref{fig_Omega}), $\Omega_r$ , $\Omega_m$ , $\Omega_\nu$  and $\Omega_\phi$ are plotted, which are the relative densities of the radiation, the pressureless matter, the mass-varying neutrinos and the  dark energy, respectively. As we can see in fig.(\ref{fig_Omega}), the universe is initially at the matter-dominated era and then at $z\simeq0.2$, the universe transits to the dark-energy-dominated era.
According to the chosen parameters (\ref{para}) and initial conditions (\ref{ICs}), relative densities in the present era, i.e. $\tau=0.92$ (corresponding to $a=1$), are obtained as  $\Omega_r=0.00009$ , $\Omega_m=0.290$ , $\Omega_\nu=0.019$  and $\Omega_\phi=0.691$ which are in the region expected by Planck 2015 data \cite{Planck}.

\section{Supplementary scalar field $\psi$ and the late time de Sitter expansion}

An interesting question is wether the model has a stable fixed point corresponding to a late time de Sitter expansion. According to (\ref{eq_set}) the evolution of energy density of the ith component is given by $\rho_i=\rho_i(a=1)a^{-3\gamma}$, where for radiation $\gamma=4/3$ and for the pressureless matter and mass-varying non-relativistic neutrinos $\gamma=1$. The fixed (critical) point of (\ref{eq_set}), at the late time, is characterized by $\dot{\phi}=0$ and $\rho_r=\rho_m=\hat{\rho}_\nu=0$. Using (\ref{rho_phi}) and (\ref{eq_set}), it is found that the de Sitter critical point is given by
\begin{eqnarray}\label{lts1}
12\epsilon H^2\phi +V_{,\phi}=0,\nonumber\\
H^2=\frac{1}{3}\frac{V(\phi)}{1-\epsilon\phi^2}.
\end{eqnarray}
By inserting the potential (\ref{potential1}) in (\ref{lts1}) we find
\begin{align}
&\phi^2_c=-\frac{\mu^2}{\epsilon\mu^2-\lambda} , \nonumber\\
&H^2_c=\frac{1}{12}\frac{\mu^4}{\epsilon\mu^2-\lambda}.
\end{align}
It is obvious that $H_c$ and $\phi_c$ cannot be real simultaneously.
Therefore, a single scalar field cannot result in a non-zero late time fixed point. Also, as it was expressed in the previous section, the single scalar field may result in turnaround and singularities.  To fix these problems, we rewrite the equations by adding a supplementary quintessence scalar field $\psi$ to the action (\ref{action}) and investigate how this additional field modifies the evolution of the system by bringing it to a stable late time de Sitter point.

Now, we extend our model in the previous section such that the new hybrid model contains these three stages:

First, when the neutrinos are relativistic, both scalar fields are in the minimum of their effective potentials ($\phi=0, \psi=0$), so the dark energy density is negligible and the universe is in the deceleration phase.

Second, when the neutrinos become non-relativistic, the effective potential of the scalar field $\phi$ becomes concave and the scalar field rolls down its effective potential. In this stage, as the dark energy density increases, the universe can enter the acceleration phase.

Third, as the scalar field $\phi$ increases, the squared effective mass of the scalar field $\psi$ becomes negative. Then the scalar field $\psi$ begins to move, too. Finally, as the other densities decrease, both fields and the Hubble parameter reach non-zero constants, so the universe enters de Sitter phase.

We assume that the scalar field $\psi$ interacts only with the scalar field $\phi$ via the potential
\begin{equation}\label{potential2}
V(\phi,\psi)=-\frac{1}{2}\mu^2\phi^2+\frac{1}{4}\lambda\phi^4-\frac{1}{2}g\phi^2\psi^2+\frac{1}{2}m^2\psi^2+\frac{1}{4}\Lambda\psi^4,\,\, (\lambda>0,\,\, \Lambda >0)
\end{equation}
The new action is given by
\begin{eqnarray}\label{action2}
S&=&\int d^4x\sqrt{-g}\left[ \frac{R}{2}-\frac{1}{2}(\partial_{\mu}\phi\partial^{\mu}\phi +\partial_{\mu}\psi\partial^{\mu}\psi +\epsilon R\phi^2) -V(\phi,\psi)\right]\nonumber\\
&&+S_\nu[\tilde{g}_{\mu\nu}]+S_m+S_r.
\end{eqnarray}
Variation of (\ref{action2}) with respect to $\psi$ yields to the equation of motion
\begin{equation}\label{EoMpsi}
\ddot{\psi}+3H\dot{\psi}+ V_{,\psi} =0.
\end{equation}
Therefore, the squared effective mass of $\psi$ derived according to (\ref{potential2}) is
\begin{equation}\label{mass_eff}
m_{eff.}^2=\frac{\partial^2 V}{{\partial \psi}^2}\Bigm|_{\psi=0}=m^2 - g\phi^2 .
\end{equation}
Also, according to the new potential, expression (\ref{mu_eff}) changes as
\begin{equation}\label{mu_eff2}
\mu^2_{eff.}=-\mu^2 - g\psi^2+\epsilon\left(\rho_m+(\rho_\nu - 3P_\nu)\right)+A^{-1}(0)A_{,\phi\phi}\Bigm|_{\phi=0}(\rho_\nu - 3P_\nu).
\end{equation}
In the first and second stages, conditions (\ref{condition1}) and (\ref{condition2}) still hold because $\psi=0$. According to (\ref{mass_eff}), because $\phi=0$ in the first stage and we want $m_{eff.}^2$ to be positive, the coefficient of $m^2$ in the potential (\ref{potential2}) must be positive. Also, we want the squared effective mass of $\psi$ to be negative in the third stage, so $g$ must be positive. Therefore, the supplementary scalar field begins to move when $\phi>\phi_c$, where $\phi_c=\sqrt{m^2/g}$.

According to action (\ref{action2}), the set of equations (\ref{eq_set}) is rewritten as
\begin{eqnarray}\label{eq_set2}
&&\dot{a}-Ha=0,\nonumber\\
&&\dot{\hat{\rho}}_\nu+3H\hat{\rho}_\nu=0,\nonumber\\
&&\dot{\rho}_m+3H\rho_m=0,\nonumber\\
&&\dot{\rho}_r+4H\rho_r=0,\nonumber\\
&&\ddot{\psi}+3H\dot{\psi} +V_{,\psi}=0,\nonumber\\
&&\ddot{\phi}+3H\dot{\phi}+\epsilon R\phi +V_{,\phi}+A_{,\phi}\hat{\rho}_\nu=0,\nonumber\\
&&-2\left(1+(6\epsilon^2-\epsilon)\phi^2\right)\dot{H}=(1-2\epsilon)\dot{\phi}^2+\dot{\psi}^2+8\epsilon H\phi\dot{\phi}+24\epsilon^2 H^2\phi^2\nonumber\\
&&+2\epsilon\phi V_{,\phi}+2\epsilon\phi A_{,\phi}\hat{\rho}_\nu+\rho_m+\frac{4}{3}\rho_r+A(\phi)\hat{\rho}_\nu.
\end{eqnarray}
As discussed at the beginning of this section, the late time solution is characterized by $\dot{\phi}=0$, $\dot{\psi}=0$ and $\rho_r=\rho_m=\hat{\rho}_\nu=0$, which according to  (\ref{fried1}), (\ref{rho_phi}) and  (\ref{eq_set2}) results in
\begin{eqnarray}\label{lts2}
&&V_{,\psi}=0,\nonumber\\
&&12\epsilon H^2\phi +V_{,\phi}=0,\nonumber\\
&&H^2=\frac{1}{3}\frac{V(\phi,\psi)}{1-\epsilon\phi^2},
\end{eqnarray}
also, the last equation of (\ref{eq_set2}) becomes $\dot{H}=0$, which determines the de Sitter regime.

Non-trivial solutions of (\ref{lts2}) are given by\footnote{There exist two other sets of solutions, where $H$ is imaginary for any choice of positive parameters.}
\begin{eqnarray}\label{sol1}
&&\phi_f^2 =\frac{m^2 g-\Lambda\mu^2-m^4\epsilon}{g^2-m^2\epsilon g+\Lambda\epsilon\mu^2-\Lambda\lambda},\nonumber\\ \nonumber\\
&&\psi_f^2 =\frac{m^2\lambda-m^2\epsilon\mu^2-g\mu^2}{g^2-m^2\epsilon g+\Lambda\epsilon\mu^2-\Lambda\lambda},\nonumber\\ \nonumber\\
&&H_f^2=\frac{1}{12}\frac{m^4\lambda-2m^2 g \mu^2+\Lambda\mu^4}{g^2-m^2\epsilon g+\Lambda\epsilon\mu^2-\Lambda\lambda}.
\end{eqnarray}
These solutions are real provided that
\begin{eqnarray}\label{condition3}
&&\frac{m^2 g-\Lambda\mu^2-m^4\epsilon}{g^2-m^2\epsilon g+\Lambda\epsilon\mu^2-\Lambda\lambda}>0,\nonumber\\ \nonumber\\
&&\frac{m^2\lambda-m^2\epsilon\mu^2-g\mu^2}{g^2-m^2\epsilon g+\Lambda\epsilon\mu^2-\Lambda\lambda}>0,\nonumber\\ \nonumber\\
&&\frac{m^4\lambda-2m^2 g \mu^2+\Lambda\mu^4}{g^2-m^2\epsilon g+\Lambda\epsilon\mu^2-\Lambda\lambda}>0.
\end{eqnarray}
In the ultimate de Sitter Universe, dark energy is dominant and we have $P_f^d=-\rho_f^d$, where $ P_f^d$ and $\rho_f^d$ are final dark energy pressure and energy density respectively given by
\begin{eqnarray}
\rho_f^d&=&V(\phi_f,\psi_f)+3\epsilon H_f^2\phi_f^2\nonumber \\
P_f^d&=&-V(\phi_f,\psi_f)-3\epsilon H_f^2\phi_f^2.
\end{eqnarray}

In order to investigate the stability of solutions (\ref{sol1}), phase space analysis may be used. To do so, we define $X^T=[x,y,u,v,z_\nu ,z_m,z_r]$\footnote{$X^T$ denotes the transpose of matrix $X$.}, where
\begin{eqnarray}\label{variable}
&&x=\frac{\dot{\phi}}{\sqrt{6}H},\,\,\ y=\phi,\nonumber \\
&&u=\frac{\dot{\psi}}{\sqrt{6}H},\,\,\ v=\psi,\nonumber \\
&&z_\nu=\frac{\sqrt{\hat{\rho}_\nu}}{\sqrt{3}H},\nonumber \\
&&z_m=\frac{\sqrt{\rho_m}}{\sqrt{3}H},\nonumber \\
&&z_r=\frac{\sqrt{\rho_r}}{\sqrt{3}H}.
\end{eqnarray}
Using these variables and the Friedmann equation (\ref{fried1}), one can find
\begin{equation}\label{fried_var}
\mathscr{V}(X_i):=\frac{V}{3H^2}=1-x^2-u^2-2\sqrt{6}\epsilon xy-\epsilon y^2-A z_\nu^2-z_m^2-z_r^2,
\end{equation}
and the fixed point corresponding to the late time solution (\ref{sol1}) is
\begin{equation}
X_f^T=[x=0,y=y_f,u=0,v=v_f,z_\nu =0,z_m=0,z_r=0],
\end{equation}
where $y_f=\phi_f$ and $v_f=\psi_f$.

According to (\ref{eq_set2}) and (\ref{fried_var}), variables (\ref{variable}) satisfy the following autonomous equations
\begin{align}\label{autonomous}
&x^\prime =E(X_i)=-\dot{\mathscr{H}}(X_i)x-3x-\epsilon\sqrt{6}\left(2+\dot{\mathscr{H}}(X_i)\right)y-\frac{3}{\sqrt{6}}A_{,y}z_\nu^2\nonumber \\
&-\frac{3}{\sqrt{6}}f(y,v)\mathscr{V}(X_i) \, ,\nonumber \\
&y^\prime =F(X_i)=\sqrt{6}x \, ,\nonumber \\
&u^\prime =G(X_i)=-\dot{\mathscr{H}}(X_i)u-3u -\frac{3}{\sqrt{6}}g(y,v)\mathscr{V}(X_i) \, ,\nonumber \\
&v^\prime =H(X_i)=\sqrt{6}u \, ,\nonumber \\
&z_{\nu}^\prime =I(X_i)=-\frac{3}{2}z_{\nu}-\dot{\mathscr{H}}(X_i)z_{\nu} \, ,\nonumber \\
&z_m^\prime =J(X_i)=-\frac{3}{2}z_m-\dot{\mathscr{H}}(X_i)z_m \, ,\nonumber \\
&z_r^\prime =K(X_i)=-2z_r-\dot{\mathscr{H}}(X_i)z_r \, ,
\end{align}
where $f(y,v):=\frac{V_{,y}}{V}$, $g(y,v):=\frac{V_{,v}}{V}$, prime denotes derivative with respect to $\ln{(a)}$ and
\begin{align}
&\dot{\mathscr{H}}(X_i):=\frac{\dot{H}}{H^2}=-\left(\frac{1}{1+(-\epsilon+6\epsilon^2)y^2}\right) \Big(3(1-2\epsilon)x^2+3u^2+4\sqrt{6}\epsilon xy\nonumber \\
&+12\epsilon^2y^2+3\epsilon y f(y,v)\mathscr{V}(X_i)+\frac{3}{2} z_m^2+2 z_r^2+\frac{3}{2} A z_\nu ^2 \Big).
\end{align}

The stability of $X_f$ may be checked as follows:\\
Setting $X \rightarrow \delta X$, one finds
\begin{equation}
\delta X^\prime =\mathcal{M} \, \delta X,
\end{equation}
where
\begin{equation}
\mathcal{M}=
\begin{bmatrix}
E_{,x} & E_{,y} & E_{,u} & E_{,v} & E_{,z_\nu} & E_{,z_m} & E_{,z_r} \\
F_{,x} & F_{,y} & F_{,u} & F_{,v} & F_{,z_\nu} & F_{,z_m} & F_{,z_r} \\
G_{,x} & G_{,y} & G_{,u} & G_{,v} & G_{,z_\nu} & G_{,z_m} & G_{,z_r} \\
H_{,x} & H_{,y} & H_{,u} & H_{,v} & H_{,z_\nu} & H_{,z_m} & H_{,z_r} \\
I_{,x} & I_{,y} & I_{,u} & I_{,v} & I_{,z_\nu} & I_{,z_m} & I_{,z_r} \\
J_{,x} & J_{,y} & J_{,u} & J_{,v} & J_{,z_\nu} & J_{,z_m} & J_{,z_r} \\
K_{,x} & K_{,y} & K_{,u} & K_{,v} & K_{,z_\nu} & K_{,z_m} & K_{,z_r}
\end{bmatrix}.
\end{equation}
The sufficient condition for (\ref{autonomous}) to have an attractor solution requires that the real part of all the eigenvalues of Matrix $\mathcal{M}$ be negative.
At $X_f$, $\mathcal{M}$ becomes
\begin{equation}
\mathcal{M}=
\begin{bmatrix}
M_{11} & M_{12} & 0 & M_{14} & 0 & 0 & 0 \\
\sqrt{6} & 0 & 0 & 0 & 0 & 0 & 0 \\
0 & M_{32} & -3 & M_{34} & 0 & 0 & 0 \\
0 & 0 & \sqrt{6} & 0 & 0 & 0 & 0 \\
0 & 0 & 0 & 0 & -\frac{3}{2} & 0 & 0 \\
0 & 0 & 0 & 0 & 0 & -\frac{3}{2} & 0 \\
0 & 0 & 0 & 0 & 0 & 0 & -2
\end{bmatrix},
\end{equation}
where
\begin{align}
&M_{11}=-3-\epsilon y_f\sqrt{6} \dot{\mathscr{H}}_{,x}(X_f)+6\epsilon y_f f(y_f,v_f) ,	\nonumber \\
&M_{12}=-\epsilon y_f \sqrt{6} \dot{\mathscr{H}}_{,y}(X_f) -2\epsilon \sqrt{6} -\frac{3}{\sqrt{6}} f_{,y}(y_f,v_f)(1-\epsilon y_f^2)+\epsilon y_f \sqrt{6} f(y_f,v_f) ,\nonumber \\
&M_{14}=-\epsilon y_f \sqrt{6} \dot{\mathscr{H}}_{,v}(X_f) -\frac{3}{\sqrt{6}} f_{,v}(y_f,v_f)(1-\epsilon y_f^2) ,\nonumber \\
&M_{32}=-\frac{3}{\sqrt{6}} g_{,y}(y_f,v_f)(1-\epsilon y_f^2) ,\nonumber \\
&M_{34}=-\frac{3}{\sqrt{6}} g_{,v}(y_f,v_f)(1-\epsilon y_f^2) .
\end{align}
Therefore, at $X_f$, matrix $\mathcal{M}$ has three eigenvalues $ -\frac{3}{2}$, $ -\frac{3}{2}$, $-2$ and four other eigenvalues which satisfy the equation below
\begin{equation}\label{xi_eq}
\xi ^4+a_3\xi ^3 +a_2 \xi ^2+a_1\xi +a_0=0 ,
\end{equation}
where
\begin{eqnarray}
&&a_3=3-M_{11},\nonumber \\
&&a_2= -\sqrt{6} M_{24}-3M_{11}-\sqrt{6} M_{12},\nonumber \\
&&a_1=\sqrt{6} M_{24} M_{11} - 3\sqrt{6} M_{12},\nonumber \\
&&a_0=6M_{12} M_{24} -6M_{14} M_{21} .
\end{eqnarray}
The necessary conditions for all the real parts of the four solutions of (\ref{xi_eq}) to be negative are
\begin{eqnarray}\label{condition4}
&&0<a_0,\nonumber \\
&&0<a_3,\nonumber \\
&&0<a_2-\frac {a_1}{a_3},\nonumber \\
&&0<a_1-\dfrac{a_3 a_0}{a_2-\dfrac {a_1}{a_3}} .
\end{eqnarray}
Obviously, these relations constrain our parameters but obtaining analytical compact form for these constraints is very complicated, if not impossible.  Although , by testing various parameters, one can see that several sets of parameters satisfy constraints (\ref{condition3}) and (\ref{condition4}). However,  in order to continue with the numerical example of the previous section, we use the considered parameters (\ref{para}) in addition to
\begin{equation}\label{para2}
g=0.5H_0^2,\,\, m=6 H_0,\,\, \Lambda=0.5 H_0^2 .
\end{equation}
One can see that these parameters result in the following solutions for equation (\ref{xi_eq})
\begin{eqnarray}
&&\xi_{1,2} = -1.92\pm1.83i ,\nonumber \\
&&\xi_{3,4} = -1.08\pm1.83i .
\end{eqnarray}
where $i=\sqrt{-1}$. Therefore, the parameters (\ref{para}) and (\ref{para2}) make all real parts of  $\mathcal{M}$'s eigenvalues negative, and according to (\ref{sol1}), the stable late time de Sitter solution is
\begin{eqnarray}\label{sol2}
&&\phi_f = 11.48 ,\nonumber \\
&&\psi_f = 7.74 ,\nonumber \\
&&H_f = 1.96 H_0 .
\end{eqnarray}
In order to complete our numerical example in the previous section with the supplementary scalar field,  we must note that according to parameters (\ref{para}) and (\ref{para2}), we have $\phi_c=6\sqrt{2}$. Also, according to the previous numerical example in $\tau=\tau_c=3.949$, $\phi=\phi_c$. Thus, the squared effective mass of the second scalar field becomes negative after $\tau=\tau_c$, which causes the scalar filed $\psi$ begin to move.
So, in order to continue our numerical evaluation, by using the values given by the previous numerical example in $\tau=\tau_c$, we consider initial conditions in $\tau=\tau_c$ as
\begin{align}\label{ICs2}
\rho_r (\tau_c)&=1.102\times10^{-9} \, H_0^2,\,\, a(\tau_c)=22.320,\,\, H(\tau_c) =1.051 \, H_0,\nonumber \\
\rho_m (\tau_c) &=7.884\times10^{-5} \, H_0^2,\,\, \phi (\tau_c) =8.485,\,\, \dot{\phi}(\tau_c)=6.029 \, H_0 , \nonumber \\
\hat{\rho}_\nu (\tau_c) &=2.623\times10^{-5} \, H_0^2,\,\, \psi (\tau_c) =0,\,\, \dot{\psi}(\tau_c)=0.01 \, H_0 .
\end{align}
So, for $0\leq \tau \leq \tau_c$ the following graphs are plotted using the set of equations (\ref{eq_set}) and initial conditions (\ref{ICs}), and for $\tau>\tau_c$, they are plotted using the set of equations (\ref{eq_set2}) and initial conditions (\ref{ICs2}).

\begin{figure}[H]
\centering
\includegraphics[width=7 cm]{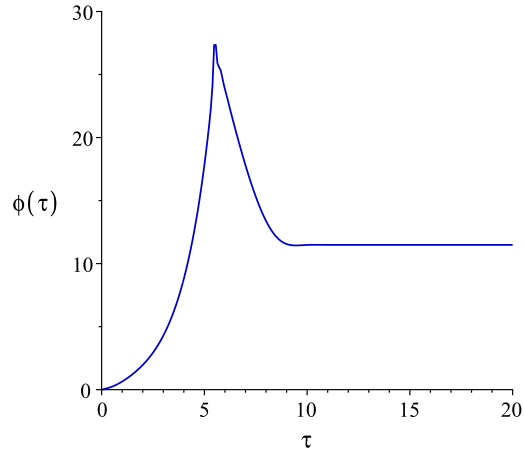}
\caption{The scalar field $\phi$ in terms of $\tau$, for the parameters (\ref{para}) and (\ref{para2}), and initial conditions (\ref{ICs}) and  (\ref{ICs2}).}
\label{fig_phi2}
\end{figure}

\begin{figure}[H]
\centering
\includegraphics[width=7 cm]{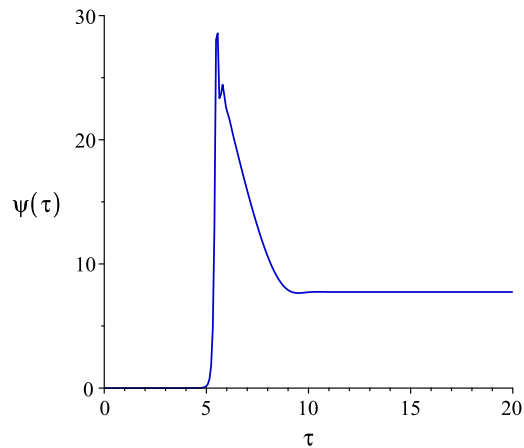}
\caption{The scalar field $\psi$ in terms of $\tau$, for the parameters (\ref{para}) and (\ref{para2}), and initial conditions (\ref{ICs}) and  (\ref{ICs2}).}
\label{fig_psi2}
\end{figure}

\begin{figure}[H]
\centering
\includegraphics[width=7 cm]{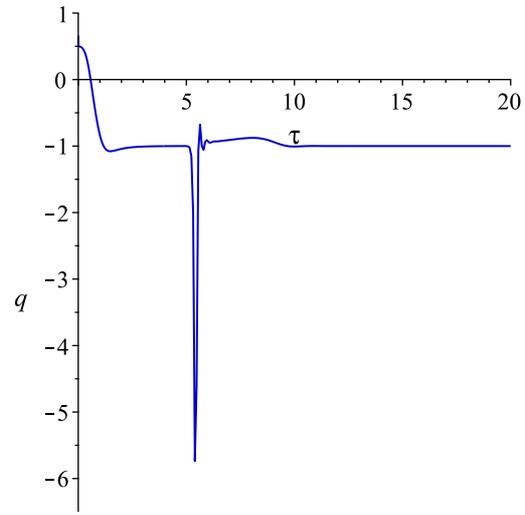}
\caption{The deceleration parameter in terms of $\tau$, for the parameters (\ref{para}) and (\ref{para2}), and initial conditions (\ref{ICs}) and  (\ref{ICs2}).}
\label{fig_q2}
\end{figure}

In fig.(\ref{fig_phi2}), the evolution of $\phi$ is plotted in terms of dimensionless time $\tau$. According to this figure,  $\phi$ grows from its insignificant initial value and eventually settles at $\phi_f=11.48$. Also,  as one can see,  with the activation of $\psi$ at $\tau=3.949$, there is no longer the singularity at $\tau=10$.
In fig.(\ref{fig_psi2}), the evolution of $\psi$ is demonstrated in terms of dimensionless time $\tau$. This evolution begins at $\tau=3.949$, and like $\phi$, $\psi$ settles at its fixed point eventually.
In fig.(\ref{fig_q2}), the deceleration parameter $q$ is plotted in terms of dimensionless time $\tau$. As one can see, in the matter-dominant era, $q=\frac{1}{2}$, and at $z=0.5$, the universe is positively accelerated and finally stays at the de Sitter point.

\newpage

\section{Summary}
Inspired by mass-varying neutrino models\cite{fardon}, and non-minimally coupled cosmological models\cite{faraoni}, we have introduced a new dynamical model to explain the origin and increment of dark energy density and also the beginning of the present cosmic acceleration. In this hybrid model, we have used two scalar fields $\phi$ and $\psi$ and a Higgs-like potential with $Z_2$ symmetry.  The first scalar field is non-minimally coupled to the Ricci scalar. This field interacts also with massive neutrinos. In one hand, this interaction makes the neutrino mass a function of the scalar field, and in the other hand relates the scalar field effective mass to the neutrinos. When the mass-varying neutrinos are relativistic, they have no interaction with the scalar field. In this stage, due to the non-minimal coupling of the quintessence to the Ricci scalar, its squared effective mass is positive; therefore, this field gets trapped in the minimum of its effective potential. As the Universe expands and its temperature decreases, the mass-varying neutrinos become non-relativistic and the shape of the quintessence effective potential changes and its squared effective mass becomes negative. Therefore, the initial stable point (with insignificant energy density) becomes unstable and the scalar field begins its evolution such that its density becomes of the same order as dark matter density. In this scenario, the coincidence problem may be related to the initial mass of mass-varying neutrinos (which determines the beginning of dark energy evolution).  We have chosen the parameters such that the ratio of densities in present time are in agreement with Planck 2015 and acceleration begins at the redshift $z\simeq0.5$.

However, this model does not have late time stable fixed point and may encounter singularities unless we introduce a second scalar field modifying the behavior of the system by bringing it to a final de Sitter stable state. The second scalar field interacts only with the first one via its potential. This field is initially trapped in the minimum of its potential. When the first scalar field passes a critical value, the squared effective mass of the second scalar field becomes negative which results in its evolution. Both fields evolve until they reach a stable attractor solution in the late time whose stability is studied in section 3. The second scalar field is necessary for such a solution.

Like the $\Lambda CDM$ model and unlike the symmetron model, the acceleration is persistent, and the universe finally experiences a de Sitter expansion. However, our model differs from the $\Lambda CDM$ model, because in this model the dark energy density is dynamical and its initial value is zero. Also, the universe may experience supper-acceleration which is an aspect of non-minimally coupled models\cite{faraoni}.

\end{document}